\documentclass[apj]{emulateapj}
\usepackage{amsmath, amsthm, amssymb}
\slugcomment{Submitted to Astrophysical Journal}

\shorttitle{AGN EXTINCTION CURVES} \shortauthors{GASKELL \&
BENKER}

\begin{document}

\title{AGN REDDENING AND ULTRAVIOLET EXTINCTION CURVES FROM HUBBLE SPACE
TELESCOPE SPECTRA}

\author{C. MARTIN GASKELL\altaffilmark{1} AND A. J. BENKER\altaffilmark{2}}
\affil{Department of Physics \& Astronomy, University of Nebraska,
\\Lincoln, NE 68588-0111} \email{gaskell@astro.as.utexas.edu}

\altaffiltext{1}{Present address: Astronomy Department, University
of Texas, Austin, TX 78712-0259}

\altaffiltext{2}{Present address: Department of Physics \&
Astronomy, University of California, Irvine, CA 92697-4575}

\begin{abstract}

We present intrinsic extinction curves for 14 AGNs.  The AGNs have
reddenings, E(B-V), of up to 0.36 mag.  The majority (13 out of 14)
of the extinction curves are not steep in the UV. Of the seven best
determined extinction curves, five have extinction curves that are
as flat as the standard Galactic curve in the optical and near UV,
but flatter in the far UV, and without the $\lambda$2175 feature.
One AGN, B3 0754+394, has a steep SMC-like extinction curve, and
another, Mrk 304, has an LMC-like extinction curve, including a
probable $\lambda$2175 bump.  The remaining seven, lower-quality,
extinction curves have overall shapes that are consistent with an
LMC-like shape or a flatter shape.  Two have possible $\lambda$2175
features, and one might be identical to the Galactic curve. The
flatter curves that predominate in our best determined extinction
curves are not as flat as the Gaskell et al.\@ (2004) extinction
curve for radio-loud AGNs. This suggests that the previous
radio-loud extinction curve might be slightly too flat in the range
$4 < 1/\lambda < 6.5$ $\mu$m$^{-1}$ because of luminosity-dependent
reddening biases in the composite spectra, but further investigation
is needed. We present a parameterized average AGN extinction curve.
Observed variations in the continuum properties of the AGNs are
inconsistent with intrinsic object-to-object variations because
observed differences are least in the far UV where changes in the
accretion disk spectrum should be greatest. We suggest that the
steepening of AGN spectra around Lyman $\alpha$ is the result of a
small amount of SMC-like dust (E(B-V) $\sim 0.03$). We find the
largest object-to-object differences in spectral shape to be in the
\ion{Fe}{2} emission of the ``small blue bump''.

\end{abstract}


\section{INTRODUCTION}

Knowing the reddening of the continua and emission lines of AGNs
is vitally important for understanding the energy-generation
mechanism of AGNs, the physical conditions of the line-emitting
gas, the differences between AGNs, and AGN demographics.

Gaskell et al.\@ (2004) showed that reddening is the major cause of
differences in optical to UV spectral energy distributions (SEDs),
and that the mean extinction in AGNs increases with decreasing
luminosity. Gaskell et al.\@ estimated mean AGN extinction curves
from composite spectra based on samples of AGNs with different mean
extinctions. These extinctions curves differ from the well known
extinction curve for dust in the solar neighborhood by lacking the
$\lambda$2175 carbon feature. The mean extinction curve for the
Small Magellanic Cloud (SMC) also lacks the $\lambda$2175 feature,
but the mean AGN extinction curves found by Gaskell et al.\@ (2004)
are considerably flatter than the SMC curve in the ultra-violet.
Czerny et al.\@ (2004) also derived a flat extinction curve with no
$\lambda$2175 bump from SDSS composite AGN colors.

The extinction curves derived by Gaskell et al.\@ (2004) and Czerny
et al.\@ (2004) refer to samples of AGNs. Using composite spectra
has the advantage of minimizing the effects of possible real
differences in the spectral-energy distributions, but the composite
spectra necessarily combined quasars of different redshifts.  For an
apparent-magnitude-limited sample the objects at higher redshift
have higher mean luminosities.  The luminosity-dependence Gaskell et
al.\@ find for the mean extinction therefore introduces the
possibility of bias in deriving a extinction curve for AGNs from
composite spectra, since the UV part of the extinction curve will
tend to come from AGNs with lower mean extinction (Willott 2005). In
this paper we avoid this bias by deriving extinction curves for
individual AGNs.

The derivation of the extinction curves is discussed in sections 2
and 3, and the curves are presented in section 4.  We consider
luminosity dependence in section 5 and in section 6 we show that
the extinction curves we derive cannot be the result of intrinsic
variation of the underlying quasar spectral energy distributions
or of varying amounts of host galaxy light. We discuss our results
in section 7.

\section{SAMPLE}

To investigate the shape of the extinction curve one needs carefully
calibrated spectra covering the whole ultraviolet and optical
spectral regions. Shang et al.\@ (2005) present spectrophotometry of
17 AGNs taken with the {\it Far Ultraviolet Spectroscopic Explorer
(FUSE)} satellite, the {\it Hubble Space Telescope (HST)}, and the
2.1-m telescope at Kitt Peak National Observatory.  The spectra
cover rest-frame wavelengths of 900 to 9000\AA \ and are thus well
suited for studying reddening.  Full observational details can be
found in Shang et al.\@ (2005).

It is important to note that though the Shang et al.\@ sample has
the necessary wavelength coverage for determining reddenings, it is
a very biased sample.  The objects were specifically chosen with the
hope that they would be bright in the far-UV (Kriss 2000). This is
going to bias the sample in favor of low-reddening AGNs.
Nonetheless, we shall see below that quite a number of the AGNs have
substantial reddening.

We show all the spectra used in Fig.\@ 1.  We do not consider the
spectral region shortwards of Lyman $\alpha$ because most of the
AGNs show absorption lines in this region (see Fig.\@ 1 of Shang et
al.\@ 2005).  All spectra have been normalized to the same zero
point at 1.8 $\mu$m$^{-1}$ and corrected for Galactic reddening. To
show the diversity of shapes better we show smoothed versions of the
spectra in Fig.\@ 2 without the offsets.  Note that the smoothing is
purely to reduce confusion for the eye in following individual
spectra in this figure, and that the smoothed spectra were not used
in the analysis.

\begin{figure} \plotone{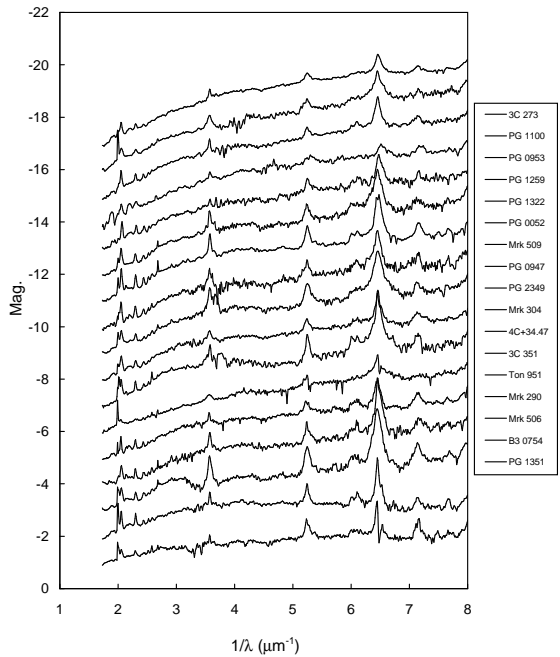} \caption{Spectra of
the FUSE/HST sample.  Spectra are shown in $F_{\lambda}$ magnitudes
normalized to 1.8 $\mu$m$^{-1}$.  Spectra have been progressively
offset by -1 mag to avoid overlap. Data from Sheng et al.\@ (2005).}
\label{fig1}
\end{figure}

\begin{figure} \plotone{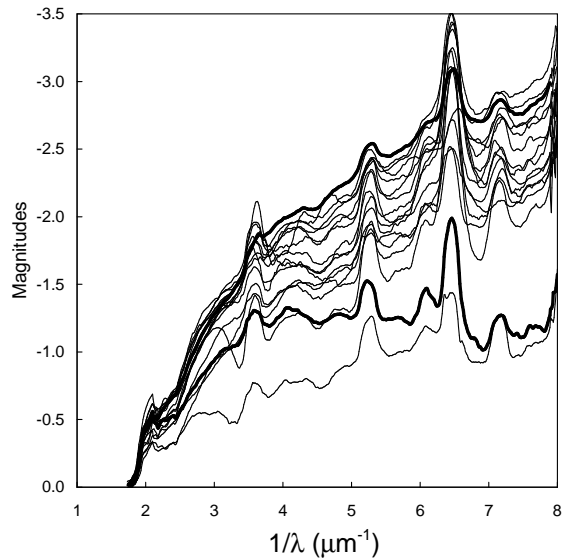} \caption{Superimposed spectra of
the FUSE/HST/optical sample.  Spectra are shown in magnitudes
normalized to 1.8 $\mu$m$^{-1}$.  The spectra have been smoothed to
make it easier for the eye to follow individual spectra.  The upper
of the two thicker lines is 3C 273 and the lower is  B3 0754+394.
Data from Sheng et al.\@ (2005).}\label{fig2}
\end{figure}

\section{DETERMINING THE EXTINCTION CURVES}

As this paper only addresses the intrinsic reddening in the AGNs,
we first removed the reddening due to dust in the Milky Way. We
used the reddening estimates of Schlegel, Finkbeiner, \& Davis
(1998) and a standard Milky Way extinction curve as parameterized
by Weingartner \& Draine (2001).

The standard method for determining extinction curves is to take
the difference in magnitudes between a more reddened object and a
less reddened object of the same intrinsic spectral shape.  This
is most commonly done for pairs of hot stars of the same spectral
class. Extinction curves are conventionally given relative to the
extinction in the V band and normalized so that the difference in
the extinction in the B band relative to the V band is one
magnitude (i.e., they are normalized to E(B-V) = 1). The scaling
factor which the observed magnitude differences relative to the V
band needs to be divided by to give the normalized extinction
curve is the difference in reddening, E(B-V), between the two
objects.

Crenshaw et al.\@ (2001) derived a extinction curve of NGC 3227 by
comparing it with NGC 4151, and Crenshaw et al.\@ (2002) did the
same for Ark 564 by comparing it with Mrk 493.  Since we have many
AGNs here, and because theory suggests the possibility of real
differences between the intrinsic spectra of different AGNs, we used
an average of several AGNs so that the extinction curves were not
just reflecting spectral peculiarities in one AGN at one epoch. We
choose as our templates the three AGNs that had the highest relative
flux from 4 to 8 $\mu$m$^{-1}$ compared with their optical fluxes:
3C 273, PG 1100+772, and PG 0953+414. Although it was not a factor
in their selection, these three AGNs happen to represent all three
main radio classes: 3C 273 is a core-dominated source, PG 1100+772
is lobe dominated, and PG 0953+414 is radio quiet.

The extinction curves in Fig.\@ 1 and Fig.\@ 2 of Gaskell et al.\@
(2004) were normalized to the B and V bands in the standard way, as
described above. However, the uncertainties in normalizing the
curves to a relatively small region in the optical are amplified
considerably by the time one gets to the UV, so we adopted a
different approach.  The differences between possible extinction
curves, such as the standard Galactic curve or the SMC curve, up to
3 $\mu$m$^{-1}$ are, for our purposes, very small, so a more
reliable procedure is to normalize the curves over the whole region
from the V-band to 3 $\mu$m$^{-1}$. Unfortunately, the region around
3 $\mu$m$^{-1}$ is in the middle of the strong broad-line region
Balmer continuum and Fe II emission making up the long-wavelength
side of the ``small blue bump'' where there are significant
object-to-object variations (see Fig.\@ 1).  We therefore moved the
baseline of our normalization to the region of 3.75 to 4.07
$\mu$m$^{-1}$.  By 4 $\mu$m$^{-1}$, known, well-determined
extinction curves begin to deviate slightly so we experimented with
slightly different normalizations covering the range from SMC
extinction curves to the Galactic extinction curve for our local
solar neighborhood.

In Fig.\@ 3 we show the differences in the separate extinction
curves of one AGN, PG 1351+640, with respect to each of the three
separate template blue AGNs.  This shows the reader the range of
uncertainty caused by using a single template AGN.  Note that the
differences between the curves would be much smaller had we
normalized between 4.5 and 6 $\mu$m$^{-1}$.  For the rest of the
paper, except where noted, we will just give extinction curves
derived with respect to the average of all three template AGNs.

\begin{figure} \plotone{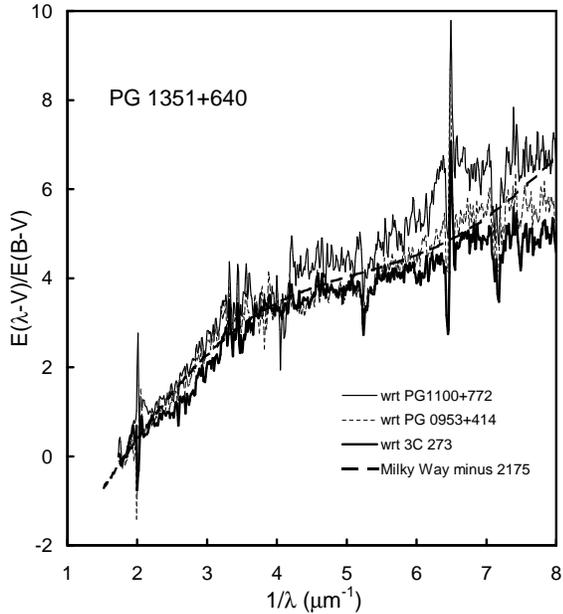} \caption{Normalized extinction curves
for PG 1351+640 derived with respect to the three template blue
AGNs.  A Milky Way extinction curve with the $\lambda$2175 carbon
feature removed is shown for comparison.}\label{fig3}
\end{figure}

In our extinction curves we have omitted the regions where there are
absorption lines. In some spectra there are problems with joining up
the UV and the optical (see Shang et al.\@ 2005), and the
signal-to-noise ratio is generally worst in these regions.  In
addition, as can be seen in Fig.\@ 3, matches are poorest around the
strong emission lines. We identified three problems in the region of
the strong UV emission lines. First, the equivalent widths of the
strong emission UV lines in AGNs decrease with increasing luminosity
-- the well-known ``Baldwin effect'' (Baldwin 1977, Baldwin et al.\@
1978) -- so to determine reddenings from the emission lines one
would have to match AGNs in luminosity. The second problem is that
strengths of the cores of the lines also vary with the so-called
``eigenvector 1'' (EV1, Boroson \& Green 1992), so to determine line
reddenings one would also have to match AGNs in EV1. The third
problem is that the profiles of the UV lines differ from object to
object because of the blueshifting of the high-ionization lines
(Gaskell 1982). All three of these effects occur to some degree in
the AGNs we consider. Since we are confining ourselves in this paper
to the issue of the shape of quasar extinction curves, in the
majority of cases we have simply deleted the parts of our extinction
curves around the major UV emission lines.

\section{EXTINCTION CURVES}

The process of normalizing the extinction curve for each AGN
produces an estimate of E(B-V). We give our E(B-V) estimates in
Table 1. Column 2 gives E(B-V)$_{\lambda2500}$, the reddening from
the normalization as described above. Column 3 gives
E(B-V)$_{\lambda1700}$, the reddening obtained by comparing the
fluxes in the 4 to 8 $\mu$m$^{-1}$ region with the optical flux,
and column 4 gives the average of the two estimates.
E(B-V)$_{\lambda2500}$ assumes that the Fe\,II emission around
$\lambda$2500 in the small blue bump is the same in all AGNs,
while E(B-V)$_{\lambda1700}$ makes the assumption that all the
extinction curves are the same shape. Our method of deriving
extinction curves automatically makes the average reddening of our
template AGNs zero.  In Table 1 we have added a small constant to
reddenings of each set to make the reddenings of the bluest
template AGN be zero. For the three AGNs with very low reddening
(see below), E(B-V)$_{\lambda2500}$ was found by comparison with
3C~273 alone.  The good agreement between E(B-V)$_{\lambda2500}$
and E(B-V)$_{\lambda1700}$ suggests that the 1-sigma uncertainty
in each E(B-V) estimate is $\pm 0.018$ mag, and that the
uncertainty in E(B-V)$_{average}$ is $\pm 0.013$ mag.

\begin{deluxetable}{lccc}
 \tablewidth{0pt}
 \tablecaption{Estimated Reddenings for AGNs}
 \tablehead{
 \colhead{Object} & \colhead{E(B-V)$_{\lambda2500}$} & \colhead{E(B-V)$_{\lambda1700}$} & \colhead{E(B-V)$_{average}$}}
 \startdata

3C 273  & 0.000 &  0.006  & 0.003\\
3C 351  & 0.186  & 0.150  & 0.168\\
4C+34.47  &  0.094  & 0.101 &  0.098\\
B3 0754+394 & 0.157  &  --- &   0.157\\
Mrk 290 & 0.166 &  0.150 &  0.158\\
Mrk 304 & 0.095  & 0.120 &  0.108\\
Mrk 506 &0.234  & 0.182 &  0.208\\
Mrk 509 & 0.043 & 0.068  & 0.056\\
PG 0052+251 & 0.032 & 0.037 &  0.035\\
PG 0947+396 &  0.095 & 0.094 &  0.095\\
PG 0953+414 & 0.025 & 0.016 &  0.021\\
PG 1100+772 & 0.013 & 0.000 &  0.007\\
PG 1259+593 & 0.014 & 0.036 &  0.025\\
PG 1322+659 & 0.064 & 0.047 &  0.055\\
PG 1351+640 & 0.366 & 0.364 &  0.365\\
PG 2349-014 & 0.077 & 0.082 &  0.080\\
Ton 951 & 0.124 &  0.155 & 0.139\\
 \enddata
\end{deluxetable}

Because the quality of derived extinction curves goes down as the
reddening decreases, we present the extinction curves in three
groups in order of decreasing reddening and increasing
uncertainty. The groups correspond to AGNs with well-determined
extinction curves, those with less certain extinction curves, and
AGNs with little reddening and hence ill-determined extinction
curves.

All extinction curves are known to be smooth in the UV, and, apart
from possible $\lambda$2175 features, we are only interested in
this smooth overall shape.  We have therefore, as discussed above,
removed the effects of mismatches in the equivalent widths of the
strong emission-lines, occasional absorption lines, and, in some
cases, the poorer signal-to-noise ratio at the junction between
the optical and {\it HST} observations.

\subsection{Well-Determined Extinction Curves}

In Fig.\@ 4 we show the extinction curves for the six reddest AGNs
in the sample.  All six lack the $\lambda$2175 feature.  Only one
AGN, B3 0754+394, has an SMC-like curve.  Four of the six have
curves that are flatter than the Galactic curve in the far UV, and
one AGN, Ton~951, is consistent with an LMC-like curve, although,
again, it is flatter in the far UV and without the $\lambda$2175
feature.

\begin{figure} \plotone{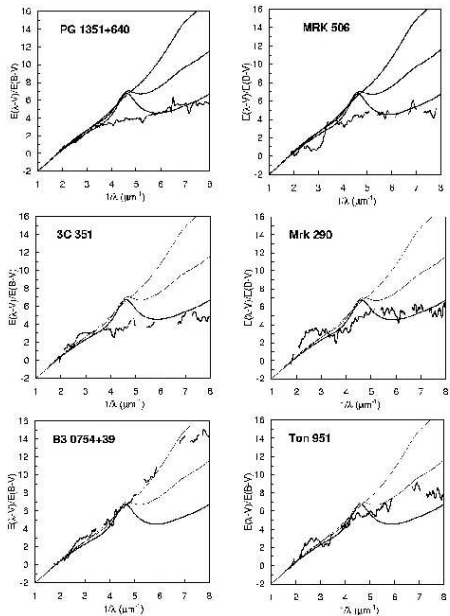} \caption{Extinction curves,
derived using the three template AGNs, for the six AGNs in the
sample with the highest extinction.  The three smooth curves in
each figure are, from top to bottom, the SMC curve, an LMC curve,
and the standard Galactic curve for dust in the solar
neighborhood.}\label{fig4}
\end{figure}

Quasars with low reddening are most sensitive to differences in the
spectral shape. These differences are largely due to the broad-line
region. Our most heavily-reddened quasar, PG 1351+640 has
significant differences in the emission-line strengths and profiles,
and also strong absorption (e.g., in C IV), but these differences
cause relatively minor differences in the extinction curve (see
Fig.\@ 4).  One of the well-known differences between AGNs is the
strength of the optical and near UV Fe II emission. In the
extinction curves one can see the effect of the varying Fe II
strength especially on the long-wavelength side of the ``small blue
bump''. This causes the extinction curves around 3$\mu$m$^{-1}$ to
be too high or too low in some objects.

\subsection{Less Certain Extinction Curves}

Fig.\@ 5 shows, in order of decreasing quality, the extinction
curves for the five next most reddened AGNs.  At lower reddening,
the spectral differences mentioned, notably the broad Fe~II emission
which is not easy to remove, become more important.  The best of the
curves, that for Mrk 304, is consistent with an LMC-like curve
including a probable $\lambda$2175 bump. The other four curves are
of poorer quality. The extinction curve of PG 1322+659 is quite
close to a Galactic extinction curve with a possible $\lambda$2175
bump, although the low extinction makes mismatching of emission
features a serious problem. PG 0947+396 is consistent with either an
LMC-like or a Galactic curve, depending on the normalization. For
both PG 0947+396 and PG 1322+659 there is probably a $\lambda$2175
bump. The two remaining curves, PG 2349-014 and 4C+34.47, are most
consistent with LMC-like curves, but the flatter shape favored by
the majority of the AGNs in Fig.\@ 4 is also possible with a
different normalization.  For 4C+34.47, in addition to the
uncertainty in the normalization caused by spectra differences
around $\lambda$2500, there is an additional uncertainty caused by a
gap between the {\it HST} spectrum and the Kitt Peak spectrum.
Inspection of Fig.\@ 1 of Shang et al.\@ (2005) suggests that the
optical spectrum could have a slightly too high flux relative to the
{\it HST} spectrum. If this is indeed the case, then the
right-hand-side of 4C+34.47 extinction curve (beyond 3.5
$\mu$m$^{-1}$) could be lowered by 2--3 magnitudes and it would then
resemble the flatter curves in Fig.\@ 4. The reddening, E(B-V), of
4C+34.47 would then be about 0.03--0.04 mag. less than the values in
Table 1. The uncertainties in the extinction curves for both
4C+34.47 and PG 2349-014 are too great to say whether $\lambda$2175
is present or absent.

\begin{figure} \plotone{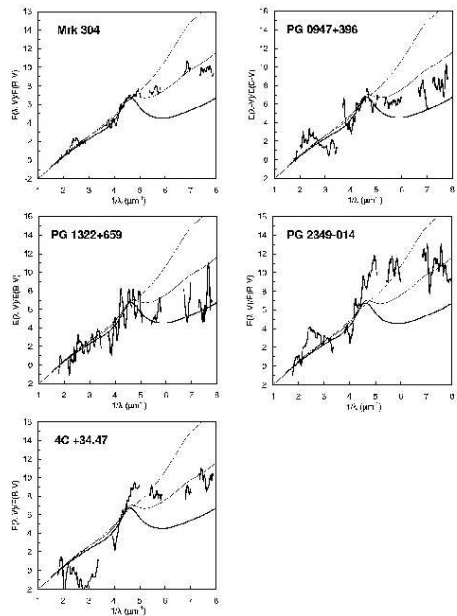} \caption{Extinction curves for
AGNs of lower extinction.  As in Fig.\@ 4, these are derived using
the three template AGNs.  The smooth curves again represent SMC,
LMC, and Galactic extinction curves.}\label{fig5}
\end{figure}

\subsection{AGNs With Low Reddening}

For the three AGNs with very low reddening we could not get
satisfactory extinction curves using all three template curves, so
we only used 3C 273.  The three extinction curves in Fig.\@ 6 are
shown for completeness and to illustrate that they favor a flatter
shape rather than a steep SMC-like shape.

\begin{figure} \plotone{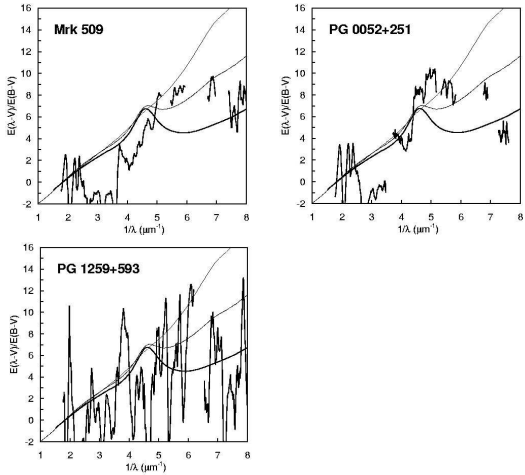} \caption{Extinction curves
for the three AGNs with the lowest extinction.  Unlike the curves
in the two previous figures, these curves were derived {\it using
only 3C 273 for a template.} The smooth curves represent SMC, LMC,
and Galactic extinction curves.  We show these curves for
completeness and to show that although the extinction curves are
noisy, (a) they favor a flatter shape rather than an SMC
extinction curve shape and (b) there is no rise in the far
UV.}\label{fig6}
\end{figure}

\subsection{The Shape of the Far UV Extinction Curve}

    One can see in Figs.\@ 4, 5 and 6 that, {\it none} of our curves
show a steep rise towards the far UV.  Only PG~0947+396 shows a
slight rise in the far UV. Even the one SMC-like curve, B3 0754+394,
seems to flatten in the far UV. Although we have not derived
extinction curves shortwards of Lyman $\alpha$, it is instructive to
look at the correlation of our reddening values with $\alpha_{FUV}$,
the far ultraviolet spectral index in a $F_{\nu} \propto
\nu^{-\alpha}$ power law, measured by Shang et al.\@ (2005) for
$\lambda \leq 1100\AA$. Their $\alpha_{UVO}$, the slope from
1200--1500$\AA$, is, as would be expected, well correlated with our
reddenings in Table 1 (correlation coefficient, $r$ = 0.87 with a
one-tailed probability of arising by chance $p < 0.001$). Since
$\alpha_{UVO}$ is dominated by reddening, one might therefore expect
$\alpha_{FUV}$ to be influenced by reddening too. As can be seen in
Fig.\@ 7, while there is a significance correlation ($p = 0.04$)
between the reddenings and $\alpha_{FUV}$, $r$ is only 0.45 (i.e.,
much less than for the correlation between $\alpha_{UVO}$ and
E(B-V).)

\begin{figure} \plotone{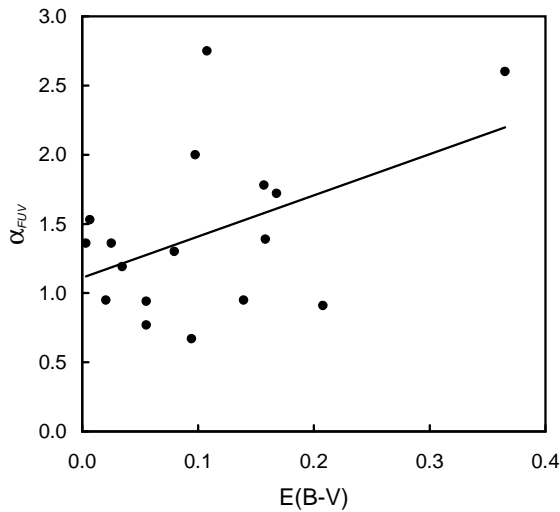} \caption{The dependence of
$\alpha_{FUV}$, the far ultraviolet spectral index, on the
reddening for the AGNs of the {\it FUSE/HST} sample.  The line is
least-squares fit.}\label{fig7}
\end{figure}

The weak dependence of $\alpha_{FUV}$ on E(B-V) supports the idea
that, unlike the Galactic curve and the SMC and LMC curves, AGN
extinction curves do not rise more steeply into the far UV. This
makes physical sense.  The rise in far-UV extinction in our Galaxy
etc. is the result of a broad spectral feature between 10 and 20 eV
that has a similar origin to that of the $\lambda$2175 feature (see
Fig.\@ 2 of Laor \& Draine 1993). Laor \& Draine (1993) consider
carbon in the form of graphite, but there is also a corresponding
spectral feature from PAH molecules (see Li \& Draine 2001). Since
the majority of our extinction curves lack obvious $\lambda$2175, we
therefore expect the far UV rise to be absent as well.  It is
interesting the one curve showing a rise in the far UV, that of
PG~0947+396, also shows a possible $\lambda$2175 feature.

Although the 10--20 eV peak goes along with the $\lambda$2175 bump,
the shape of this 10--20 eV feature is much more sensitive to the
form the carbon is in and the grain size distribution than the
$\lambda$2175 feature. Inspection of the {\it FUSE} composite AGN
spectrum of Scott et al.\@ (2004) shows that the sharp peak in the
theoretical extinction curve of Weingartner \& Draine (2001) due to
small graphite grains (see Fig.\@ 9 of Draine 2003) is certainly
absent.

The apparently flat far UV extinction curve of AGNs does raise one
puzzling issue with the far UV: what is the cause of the turndown in
the spectral shape in the far UV? As Shang et al.\@ (2005) show,
$\alpha_{FUV}$ is steeper than $\alpha_{UVO}$ by $\sim 1$.  Binette
et al.\@ (2005) suggest that this is due to the presence of
nanodiamonds in AGN dust.  We suggest, however, that perhaps a more
natural way of producing this steepening would be by having
extinction of E(B-V) $\sim 0.03$ mag caused by intervening dust with
a Galactic, LMC, or SMC extinction curve, as Shang et al.\@ point
out. In order not to show up in our extinction curves this dust with
a steeper far UV extinction curve would have to be independent of
the nuclear dust in the AGN. In agreement with this, our three
template AGNs (the rightmost points in Fig.\@ 7) show average
$\alpha_{FUV}$ slopes.

\subsection{A Mean Extinction Curve}

It has long been common practice to deredden AGNs with a standard
Galactic extinction curve. As we have shown above however, AGN
reddening curves differ from the standard Galactic extinction curve
by lacking the $\lambda$2175 feature and by being flatter in the far
UV. In order to provide a better typical AGN extinction curve for
de-reddening AGNs we have constructed a mean curve from the
extinction curves of the AGNs with the highest reddening, but
excluding the SMC-like curve of B3 0754+394.  The mean curve was
constructed as follows: firstly fourth or fifth order polynomials
were fit to the extinction curves of each of the five AGNs through
the regions not heavily effected by strong emission lines, and then
an unweighted average was then taken of these polynomials. This
average extinction curve is shown in Fig.\@ 8 along with the errors
in the mean curve. These errors are a minimum around the two regions
in which the five input curves were normalized (see section 3).  We
also show the Galactic extinction curve for the solar neighborhood,
and the Galactic extinction curve with the Drude profile of the
$\lambda$2175 feature removed.  As can be seen from Fig 8, the
average AGN extinction curve is in excellent agreement with the
Galactic curve from the optical up to $\sim 4 \mu$m$^{-1}$ and
possibly to 6 or 7 $\mu$m$^{-1}$, but is significantly flatter
beyond 7 $\mu$m$^{-1}$

\begin{figure} \plotone{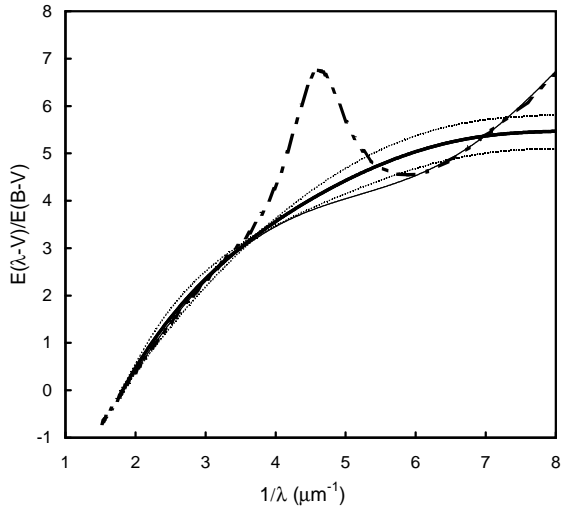} \caption{The average extinction
curve for the five AGNs with greatest reddening (see text).  The
mean AGN curve is shown as a bold line with 1-sigma error bars for
the error in the mean shown as dotted lines on either side of it.
The solid dashed-dotted curve is the Galactic extinction curve for
the solar neighborhood, and the thin line is the Galactic curve
with the $\lambda$2175 feature removed.}\label{fig8}
\end{figure}

For convenience, the mean extinction curve can be represented by a
fifth-order polynomial in $x = \lambda^{-1}$ in $\mu m^{-1}$, valid
over the range $1.5 < x < 8$ (i.e., from H$\alpha$ $\lambda$6564 to
Lyman $\alpha$ $\lambda$ 1216).

\begin{align*}
E(\lambda-V)/E(B-V) = 0.000843x^5 - 0.02496x^4 \\ + 0.2919x^3
-1.815x^2 + 6.83x - 7.92
\end{align*}

This equation supersedes the relationships given in the Appendix of
Gaskell et al.\@ (2004).  For longer wavelengths a standard Galactic
extinction curve can be used.

Although the parametrization in Eq.(1) gives a more appropriate
extinction curve than assuming that the AGN extinction curve is the
same as the Galactic curve, researchers are cautioned that
individual AGN extinction curves vary, as can readily be seen in
Figs.\@ 4 -- 6., so one should, if possible, determine the
extinction curve directly for each individual AGN they are studying.

\section{LUMINOSITY DEPENDENCE}

In Fig.\@ 9 we show the luminosity dependence of our reddenings for
the {\it FUSE/HST} sample compared with the luminosity--mean
reddening relationship from Gaskell et al.\@ (2004).  As would be
expected, given the bias in the {\it FUSE/HST} sample towards
UV-bright objects, the reddenings of the AGNs at a given luminosity
mostly lie below the typical values found by Gaskell et al.\@  It
can be seen, however, that the two most reddened AGNs for their
luminosities have reddenings quite typical of the reddenings in the
samples in Fig.\@ 5 of Gaskell et al.\@ (2004).

\begin{figure} \plotone{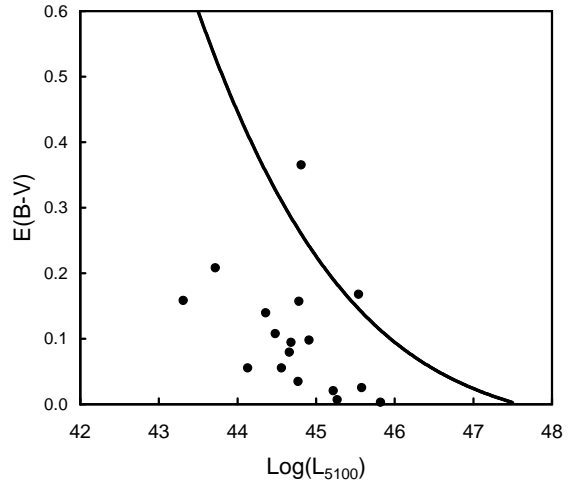} \caption{The luminosity
dependence of the reddenings of the AGNs in the {\it FUSE/HST}
sample.  The solid line is from Fig.\@ 5 of Gaskell et al.\@
$L_{5100}$ is $\lambda$$L_{\lambda}$ at $\lambda$5100 in ergs
s$^{-1}$ from Shang et al.\@ (2005).}\label{fig9}
\end{figure}

\section{CAN REDDENING BE IMITATED BY DIFFERENCES IN SPECTRAL-ENERGY
DISTRIBUTIONS?}

With some notable exceptions (e.g., Carleton et al.\@ 1987) it has
long been widely considered that apparent differences in AGN
spectral energy distributions (SEDs) are caused by intrinsic
differences. For example, Yip et al.\@ (2004) concluded from their
eigenspectral analysis of a very large sample of SDSS quasars that
the spectral differences are primarily due to intrinsic differences
in the AGN spectral energy distribution and to varying amounts of
host galaxy light. In this section we quantitatively explore these
effects and show that they cannot reproduce our observed reddening
curves.

Winkler et al.\@ (1992) find no evidence that the {\it optical}
spectrum changes its shape in an individual AGN as it varies, but
Barr, Willis \& Wilson (1983), and Clavel et al.\@ (1991) observe a
hardening of the UV spectra of some AGNs when they brighten.  On the
other hand it is notable that in the case of Fairall 9 the UV
continuum slope stayed the same at $\alpha_{UV} = 0.46 \pm 0.11$
while the UV flux changed by a factor of over 20 (Clavel et al.\@
1989).  We will model possible spectral shape changes in two ways:
as a varying ``big blue bump'' contribution, and as a change in the
relative contribution of two power-laws.

\subsection{The Effect of Different Big Blue-bump Contributions}

Krolik et al.\@ (1991) suggest that the luminosity of the ``blue
bump'' varies while an underlying $\nu^{-1}$ power-law remains
constant. Such variation in an individual object is inconsistent
with the optical and IR observations of Winkler et al.\@ (1992), and
it is easy to demonstrate that such a variation {\it between}
objects will {\it not} explain the extinction curves we see.

In the ``power law + blue bump'' picture, the blue bump is of a
thermal origin and has an exponential Wien cutoff.  We approximated
the bump as a single black body and added an arbitrary $F_{\nu}
\propto \nu^{-1}$ power law. We varied the temperature of the black
body, and derived pseudo-extinction curves from the resulting
spectra. These curves are normalized to E(B-V) = 1 in the standard
manner . We show our results in Fig.\@ 10 along with Milky Way and
SMC extinction curves.

\begin{figure} \plotone{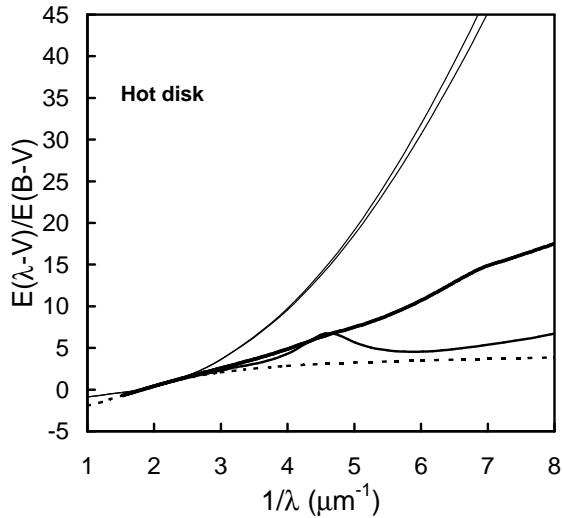} \caption{Psuedo-extinction curves
(thin curves) resulting from a change in the temperature of a hot
disk compared with Milky Way (lower thick curve) and SMC (upper
thick curve) extinction curves. The top thin curve comes from
thermal components of temperatures 300,000K and 200,000K that are
28\% and 45\% stronger than the power law at Lyman $\alpha$. The
curve just below it is derived from components with similar
relative strengths and temperatures of 100,000K and 50,000K.  The
dotted curve is derived from components of temperatures 300,000K
and 200,000K that are 50 times and 80 times stronger than the
power law at Lyman $\alpha$.}\label{fig10}
\end{figure}

The SED of AGNs probably peaks in the EUV/soft X-ray region of the
spectrum (see Gaskell, Klimek \& Nazarova 2007). This would
correspond to temperatures of 100,000 to 300,000 K. We find that the
dominant parameter affecting the pseudo-reddening curves is not the
temperature or the slope of the arbitrary power law, but the ratio
of the thermal component to the arbitrary power law. As can be seen
from Fig.\@ 10 the choice of temperature range has little effect on
the shape of the pseudo-extinction curves, but varying the ratio of
the thermal component to the arbitrary power law has a large effect.
The higher the ratio of the thermal component to the arbitrary power
law, the flatter the pseudo-extinction curve.

Although it might appear from Fig.\@ 10 that it is easy to produce
smooth extinction curves resembling those we have found, this is
very misleading. In reality there are two major problems. First, one
cannot produce any arbitrary reddening. The largest pseudo reddening
produced in Fig.\@ 10 is only E(B-V) = 0.076. Worse still, the
pseudo reddening is tied to the shape of the extinction curve. The
second major problem is that the shapes of the spectra being
compared do not resemble real AGNs.  In the {\it HST} composite of
Zheng et al.\@ (1997) the power-law index, $\alpha$, of the spectral
region around Lyman $\alpha$ is 1.72.  From the {\it FUSE} composite
of Scott et al.\@ (2004) $\alpha$ = 0.52.  For the cases where our
pseudo-extinction curves resemble the AGN extinction curves derived
above, our thermal component plus arbitrary power law model gives
steeply {\it rising} spectra with $\alpha \approx$ -1.7.  Such
spectra are never observed.

If we try cooler thermal components with temperatures of around
30,000 K (see Fig.\@ 11) we can only produce SMC- or LMC-like
curves. Again, the pseudo-extinction curves depend primarily on the
strength of the thermal component relative to the arbitrary power
law and are insensitive to the choice of temperatures.  It is easier
to mimic more significant reddenings.   However, the spectral shape
problem remains.  The 40,000K black body plus power law produces
gives a rising spectrum of $\alpha \approx$ -0.5 in the spectral
region around Lyman $\alpha$.  This is again at variance with
observations.

\begin{figure} \plotone{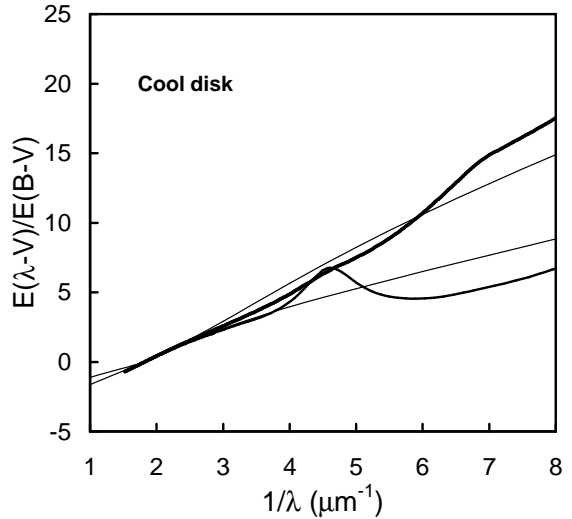} \caption{Psuedo-extinction curves
(thin curves) resulting from a change in the temperature of a cool
disk. The thick lines are Milky Way and SMC curves as in Fig.\@ 10.
Both of the thin curves come from comparing thermal components of
temperatures 40,000K and 20,000K.  For the upper curve the thermal
components are 37\% and nine times stronger than the power law at
Lyman $\alpha$.  For the lower thin curve they are 5 times and 37
times stronger than the power law at Lyman $\alpha$. The top and
bottom curves correspond to E(B-V) = 0.098 and 0.175 mag
respectively.}\label{fig11}
\end{figure}

\subsection{Two Power Laws?}

We also considered an arbitrary mixture of power-laws.  The overall
optical to X-ray spectral shape of most AGNs can be described with
an $\alpha$ = 1.7 power law.  Winkler et al.\@ (1992) present the
shapes of the variable continua for many AGNs. We took the bluest of
these to represent the unreddened variable continuum and represent
this with an $\alpha$ = -0.2 power law. In Fig.\@ 12 we show the
pseudo-extinction curves obtained by adding increasing amounts of
the $\alpha$ = -0.2 power law to the $\alpha$ = 1.7 power law.  The
precise choice of spectral indices of the two components is not
important.  As can be seen from Fig.\@ 12, it is possible to
reproduce SMC- or LMC-like curves and somewhat higher apparent
reddenings, but the same problems exist as with the thermal
components in Fig.\@ 11: the shape of the extinction curve is
coupled to the degree of reddening (see caption to Fig.\@ 11), it is
not possible to get very high reddenings, and for all but the most
modest reddenings, the shapes of the bluest spectra are again much
steeper than is observed.  In Fig.\@ 13 we show the continuum
spectral indices around Lyman $\alpha$ for the bluest spectra as a
function of the pseudo reddening.

\begin{figure} \plotone{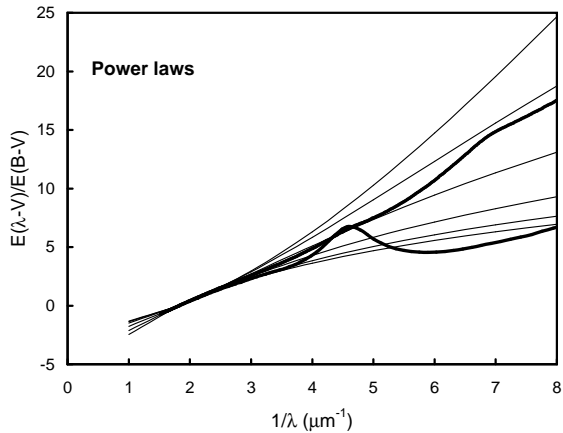} \caption{Psuedo-extinction curves
(thin curves) resulting from adding varying amounts of an $\alpha$ =
-0.2 power law to an $\alpha$ = 1.7 power law. The thick lines are
Milky Way and SMC curves as in Figs.\@ 10 and 11.  The thin lines
correspond to effective reddenings of E(B-V) = 0.02, 0.05, 0.13,
0.25, 0.36, \& 0.43 magnitudes.}\label{fig12}
\end{figure}

\begin{figure} \plotone{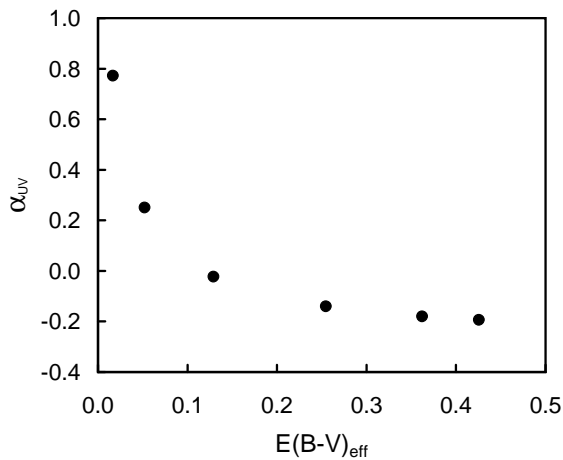} \caption{Continuum spectral index
in the Lyman $\alpha$ spectral region for the bluest AGNs as a
function of the pseudo reddening, E(B-V)$_{eff}$.}\label{fig13}
\end{figure}

\subsection{The Effect of Differing Host Galaxy Contributions}

The final effect we consider is the idea that the differences in
spectral energy distributions are due to changing host galaxy
contributions (Yip et al.\@ 2004).  We used a simple cool black body
to approximate the effect of a changing host galaxy stellar
population and approximated the quasar continuum as a $\nu^{-1}$
power law.  As is to be expected, all the changes are in the long
wavelength end of the spectrum, and if the changes in spectral shape
are interpreted as a pseudo-extinction curve (see Fig.\@ 14), the UV
part of the curve bears no resemblance to a reddening curve at all.

\begin{figure} \plotone{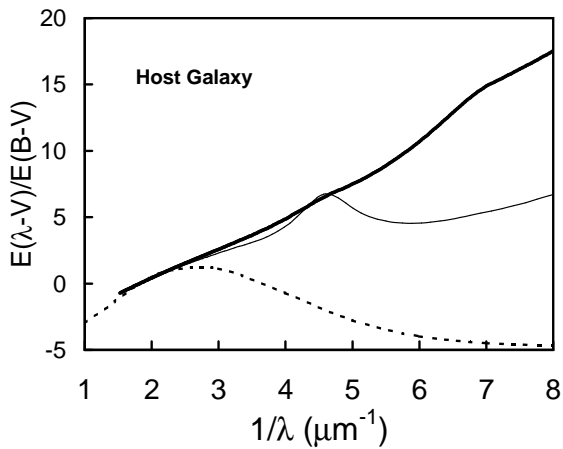} \caption{Similar to Figs.\@
11 and 12, but showing the effect of a cooler black body changing
its temperature from 4000 to 5000 K (i.e., the range of effective
temperature of an old stellar population), and its contribution to
the flux at 5500$\AA$ from 19\% to 72\% compared with a $\alpha =
-1$ power law.}\label{fig14}
\end{figure}

From the analysis presented in this section we conclude that
changes in the continuum shape from AGN to AGN, either because of
changes in the AGN itself or in the degree of host galaxy
contamination, are {\it not} the cause of the extinction curves we
have derived. This conclusion is also supported by the similarity
of broad-line and continuum reddenings, and reddenings derived
from X-rays (see discussion in section 7.6).  A positive
conclusion that can be drawn from the analysis of sections 6.1 and
6.2, however, is that because changes in continuum shape do mimic
reddening to some degree, small changes in the shape of the
continua from object to object, if such changes exist, will not be
significantly distorting the shapes we derive.

\section{DISCUSSION}

\subsection{The Fraction of Steep Curves}

Richards et al.\@ (2003) only consider the possibility of AGNs being
reddening by SMC-like dust, and Willott (2005) also argues for AGNs
being reddened by SMC-like dust, but in our study we find only one
clear case of a steep, SMC-like reddening curve, B3 0754+394. We
believe the paucity of SMC-like curves is supported by other
studies.  Gaskell et al.\@ (2004) found quite flat extinction in the
UV for the samples of AGNs they considered, especially the
radio-loud AGNs. Crenshaw et al.\@ (2001) found a clearly steep,
SMC-like extinction curve for the radio-quiet Seyfert NGC 3227
(E(B-V) $\approx 0.180$ compared with NGC~4151). However, the
extinction curve Crenshaw et al.\@ (2002) found for Ark 564 is
similar to a number of those in Figs.\@ 4 and 5.  Gaskell, Klimek,
\& Nazarova (2007) find an extinction curve for NGC~5548 similar to
our average curve presented here.

Richards et al.\@ (2003) only consider SMC-like reddening curves.
Gaskell et al.\@ (2004) have pointed out that the statement by
Richards et al.\@ that attempting to explain their large
color-segregated composites with reddening ``results in good matches
at both 1700 and 4040 \AA$ $ but overpredicts the flux between these
two wavelengths and underpredicts the flux shortward of C IV'' shows
that the reddening curve for the AGNs they consider must in fact be
flatter than an SMC-like curve.

As noted in section 2, the FUSE/HST/optical sample of Shang et al.\@
(2005) is biased towards AGNs with low reddening.  More
specifically, it is biased towards AGNs with low {\it ultraviolet}
reddening.  It will therefore be particularly biased against AGNs
which are reddened by SMC-like extinction curves.  We therefore
predict that a less biased sample would not only have more AGNs with
higher reddening, but could also have more with steeper (e.g.,
SMC-like) extinction curves.  However, the small spread in the UV
slope of AGNs (Cheng, Gaskell, \& Koratkar 1991) suggests that
steep, SMC-like curves are not common.

\subsection{The $\lambda$2175 Feature}

The $\lambda$2175 feature has long been known to be weak in AGNs
(McKee \& Petrosian 1974). Of the six highest quality extinction
curves presented here (see Fig.\@ 4), none show the $\lambda$2175
feature, while there is an indication that three of the next highest
quality five extinction curves might show it. Nonetheless we believe
that the $\lambda$2175 feature is relatively rare in AGN extinction
curves because it is completely absent in the extinction curves
presented by Czerny et al.\@ (2004) and Gaskell et al.\@ (2004)
based on composite spectra. Although, as noted in the introduction,
luminosity biases can make extinction curves from composite spectra
(e.g., those of Gaskell et al.\@ 2004, and Czerny et al.\@ 2004)
flatter in the UV than they ought to be (Willott 2005), one thing
this will {\it not} do is affect the strength of $\lambda$2175.

\subsection{Emission-line Reddenings}

Emission-line reddenings can be estimated from known line ratios.
The \ion{He}{2} $\lambda$1640/$\lambda$4686 and \ion{O}{1}
$\lambda$1304/$\lambda$8446 ratios are particularly useful. For Mrk
290 the redshift is low enough that \ion{O}{1} $\lambda$8446 was
observed, and the optical \ion{Fe}{2} emission is weak enough that
\ion{He}{2} $\lambda$4686 can be measured.  Relative line strengths
were estimated by profile fitting.  We fit the \ion{He}{2} lines
with the \ion{C}{4} profile and the \ion{O}{1} lines with the
H$\alpha$ profile. We estimate that the line ratios are uncertain by
up to about 50\%. This is because the lines are weak and determining
the appropriate underlying continuum in the presence of other weak
lines and the wings of stronger lines is difficult. For Mrk 290 we
estimate the \ion{He}{2} $\lambda$1640/$\lambda$4686 ratio to be
3.8. If the real ratio is 8.3, as suggested by photoionization
models, then our average extinction curve implies a total reddening,
E(B-V) of 0.20, or an intrinsic reddening of 0.18 after allowance
for Galactic reddening of 0.016 mag. We estimate the \ion{O}{1}
$\lambda$1304/$\lambda$8446 ratio to be 3.1 compared with the
theoretical value of 6.5.  This implies an intrinsic reddening of
0.17 mag.

These two reddening estimates are in agreement with our estimate of
0.16 mag from the continuum shape (i.e., the value in Table 1). The
errors in the reddenings estimated from the \ion{O}{1} and
\ion{He}{2} ratios could be up to $\pm 0.1$ mag, but the agreement
provides support for the differences in observed continuum shapes
being due to reddening. Further careful study of these line ratios
in more AGNs would be very worthwhile to check what the zero-point
of the reddening is (i.e., what the reddening is for our template
AGNs).

\subsection{The Shape of the Spectral-Energy Distribution}

Emission lines provide additional support for our contention that
apparent differences in the optical-UV spectral shape are first and
foremost due to reddening.  We have shown that reddening, and hence
the apparent optical to UV spectral index, is luminosity dependent
(see Gaskell et al.\@ 2004 and section 9 above). There is no doubt
that broad emission lines are produced by photoionization resulting
from an ionizing continuum that is an extrapolation of the observed
UV-optical continuum.  If the UV-optical continuum really were
steeper at lower luminosities, rather than being reddened more, then
there would be far fewer ionizing photons on average in
lower-luminosity AGNs.  This would result in much lower
emission-line equivalent widths in lower-luminosity AGNs, which is
completely the opposite of the Baldwin effect.

With the possible exception of the far UV (see section 4.4), in
the spectral region we consider, the only spectral differences we
detect that are clearly not due to reddening are due to differing
broad-line region properties.  These include differing strengths
and widths of the major emission lines, and differences in the
``small blue bump'' due to Balmer continuum and Fe~II emission.

\subsection{Implications of Flat Versus Steep Reddening Curves}

The shape of the reddening curve strongly affects the estimated
extinction.  For a given extinction at $\lambda$1640, say, a flat
curve such as that in Fig.\@ 8 implies an E(B-V) about three times
greater than what an SMC curve would imply.  This means that by
assuming an SMC-like curve {\it a priori}, Richards et al.\@ (2003)
are underestimating the reddening of SDSS AGNs (e.g., in their
Fig.\@ 6) by a factor of 3--5 at higher redshifts.

The larger grain sizes producing the flatter cure could also mean
that the ratio, $R$, of V-band extinction, $A_V$, to E(B-V) is
greater than the value of 3.1 commonly found in the local ISM.
Gaskell et al.\@ (2004) suggest that $R \approx 5$ is appropriate
for their flat extinction curve.

Taken together, the increased E(B-V) and, to a lesser extent, the
possibility of an increased $R$, means that $A_V$ is a lot greater
with a flatter extinction curve than with an SMC-like curve.  This
has important implications for AGN energetics and demographics.

\subsection{The Amount of Reddening of AGNs}

De Zotti \& Gaskell (1985) find, from considerations of broad-band
optical colors and Balmer decrements as a function of orientation,
that the typical reddening for a large sample of optically-selected
Seyfert 1 galaxies is $\sim 0.30$ mag. Winkler et al.\@ (1992)
derive reddenings for a large sample of southern Seyferts from
Balmer decrements, the X-ray to H$\beta$ ratio, and variability
colors.  The mean reddenings from their three methods are E(B-V) =
0.32, 0.41, and 0.29.  From a variety of broad-line ratios Ward \&
Morris (1984) get E(B-V) = 0.27 mag. for NGC 3783. Estimates of the
reddening from the UV/optical continuum shapes of individual AGNs
are consistent with these reddenings. For NGC 3227 Crenshaw et al.\@
(2001) get E(B-V) = 0.18 relative to NGC 4151, and for Ark 564
Crenshaw et al.\@ (2002) get E(B-V) = 0.14 compared with Mrk 493.
For NGC 5548 Gaskell et al.\@ (2005) get E(B-V) = 0.20 compared with
3C 273. All of these AGNs were optically selected. Considering all
of these reddening estimates we see that, as expected, the
reddenings of the FUSE/HST/optical sample are lower than average for
AGNs.

When one looks at complete samples of AGNs selected by radio
emission, the reddenings are even greater. For the four increasingly
reddened Molonglo samples of Baker \& Hunstead (1995) Gaskell et
al.\@ (2004) get mean reddenings of E(B-V) = 0.29, 0.34, 0.71, and
0.98 mag. Thus the Baker \& Hunstead samples are reddened more than
typical optically-selected Seyfert galaxies, and the least reddened
of the Molonglo samples is more reddened than the FUSE/HST/optical
sample of Shang et al.\@ (2005).

From all these diverse reddening estimates it is difficult to avoid
the conclusion that reddening must have a strong effect on the
apparent spectral shapes of AGNs.  Additional, independent support
for steeper slopes being a consequence of dust comes from the
correlation between spectral index and the strength of associate
absorption (Wills, Shang, \& Yuan 2000, Baker et al.\@ 2002).

\section{Conclusions}

In agreement with Gaskell et al.\@ (2004) we argue that the
UV-optical continuum shape is profoundly affected by reddening for
all but the bluest AGNs.  We believe that all apparent differences
in the UV-optical continua of AGNs are primarily due to differences
in extinction and in the forms of the extinction curves.  Until the
effects of reddening are carefully allowed for it is not possible to
investigate possible remaining differences in the spectral energy
distributions of AGNs as a function of such things as luminosity,
black-hole mass, accretion rate, or radio power. At present all AGNs
have optical-UV continuous spectra that appear to be consistent with
a single shape after allowance for the effects of reddening and the
small blue bump.  Even in a sample, such at the {\it HST/FUSE}
sample, biased towards low reddening, most of the objects show
detectable reddening.

Our results support the interpretation of Shang et al.\@ (2003) that
the second eigenvector in their spectral principal component
analysis is due to dust reddening, and disagree with the
interpretation of Yip et al.\@ (2004) that the main continuum
eigenvectors are due to changing host galaxy contributions and
variations in the intrinsic spectral index.

We find rising SMC-like reddening curves to be rare in the
relatively low reddening {\it HST/FUSE} sample we consider and
believe that they are not the most common reddening curve for
AGNs. Flatter reddening curves imply that the visual extinction of
AGNs is greater than if the reddening curves were SMC-like.

We confirm that the $\lambda$2175 extinction feature is mostly
absent in AGN extinction curves. The rise into the far UV seen in
the Galactic, LMC, and SMC reddening curves seems also to be
mostly absent.  These differences imply that the dust has been
modified by the AGN environment.

Our new mean AGN extinction curve, while flat in the far UV
(1/$\lambda > 6.5$ $\mu$m$^{-1}$) is steeper in the range $3.5 <
1/\lambda < 6.5$ $\mu$m$^{-1}$ than the Gaskell et al.\@ (2004)
extinction curve from composite spectra of radio-loud AGNs. Although
the objects are not the same, the difference suggests that
luminosity-dependent extinction (Willott 2005) might have caused the
Gaskell et al.\@ curve to be too flat in the UV.  In disagreement
with Willott (2005), however, we do not find the majority of AGN
extinction curves to be SMC-like.

\acknowledgments

We are grateful to Zhaohui Shang and Mike Brotherton for making the
UV and optical spectra available in a convenient form and for useful
discussions. We would also like to acknowledge helpful discussions
about AGN extinction and related issues with Ski Antonucci and Liz
Klimek, and discussion of dust properties with Bruce Draine and Joe
Weingartner. We are grateful to Daniel Gaskell for assistance with
writing software.  This research has been supported by the
University of Nebraska UCARE program, by the National Science
Foundation through grant AST 03-07912, and by the Space Telescope
Science Institute through grant AR-09926.01.



\end{document}